*Article*

# Chaotic Time-Delay Signature Suppression and Entropy Growth Enhancement Using Frequency-Band Extractor

Yanqiang Guo [1,2], Tong Liu [1], Tong Zhao [1], Haojie Zhang [1] and Xiaomin Guo [1,*]

1. Key Laboratory of Advanced Transducers and Intelligent Control System, Ministry of Education, College of Physics and Optoelectronics, Taiyuan University of Technology, Taiyuan 030024, China; guoyanqiang@tyut.edu.cn (Y.G.); liutong0912@163.com (T.L.); zhaotong.tyut@outlook.com (T.Z.); tyut_zhanghaojie@163.com (H.Z.)
2. State Key Laboratory of Quantum Optics and Quantum Optics Devices, Shanxi University, Taiyuan 030006, China
* Correspondence: guoxiaomin@tyut.edu.cn

**Abstract:** By frequency-band extracting, we experimentally and theoretically investigate time-delay signature (TDS) suppression and entropy growth enhancement of a chaotic optical-feedback semiconductor laser under different injection currents and feedback strengths. The TDS and entropy growth are quantified by the peak value of autocorrelation function and the difference of permutation entropy at the feedback delay time. At the optimal extracting bandwidth, the measured TDS is suppressed up to 96% compared to the original chaos, and the entropy growth is higher than the noise-dominated threshold indicating that the dynamical process is noisy. The effects of extracting bandwidth and radio frequencies on the TDS and entropy growth are also clarified experimentally and theoretically. The experimental results are in good agreements with the theoretical results. The skewness of the laser intensity distribution is effectively improved to 0.001 with the optimal extracting bandwidth. This technique provides a promising tool to extract randomness and prepare desired entropy sources for chaotic secure communication and random number generation.

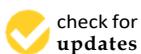

## 1. Introduction

Semiconductor laser subject to optical feedback is an all-optical and versatile structure to produce high-dimensional chaotic signals, which provides excellent test beds for the study of nonlinear dynamics [1–3]. Intrinsic quantum noise is rapidly and nonlinearly amplified by the dynamics, resulting in complexity and unpredictability of macroscopic outputs after long time [4,5]. With the deep study of optical-feedback semiconductor lasers and their complex outputs, the chaotic lasers have been applied to secure communication [6–8], high-precision ranging radar [9], optical time-domain reflectometer [10], optical fiber sensing [11] and physical random number [12–14]. However, the time delay signature (TDS) introduced by external optical feedback and relaxation oscillation period deteriorate the performance in applications of secure communication [1], chaotic radar [9] and random number generation [14]. Only a part of the chaotic bandwidth can be utilized effectively and it is difficult to rapidly access the physical process of entropy evolution.

Up to now, many schemes of TDS suppression for chaotic laser have been reported, which are mainly divided into two types of approaches. The first type for TDS suppression relies on different alternative mechanisms, such as mutually coupled system [15–17], frequency-detuned grating feedback [18], fiber Bragg grating feedback [19], random distributed feedback [20], analog-digital hybrid feedback [21], phase-modulated feedback [22–24], and parameter mismatch [25]. In the second type, the TDS is suppressed by post processing, like exclusive-OR operation [12], the m least significant bits selection [26], and optical-electrical heterodyne [13,27]. It is worth noting that the entropy of chaos is enhanced while the TDS is suppressed [28]. Due to the advantages of permutation entropy (PE) including easy measurement and strong robustness [29], the PE value at the delay time of external cavity has been applied to experimentally quantify the complexity and weak-strong chaos transition [30–32]. The effect of noise on the PE of nonlinear dynamical systems is also analyzed [33]. However, the above schemes of TDS suppression do not sufficiently avoid uneven spectrum, limited bandwidth and low energy utilization around relaxation oscillation period, which seriously limits information transmission speed and key distribution rate of chaotic optical communication. Despite calculating the PE of each embedding dimension, it is also difficult to fast assess whether the real chaotic dynamics is dominated by noise or not.

In this paper, the TDS suppression and entropy growth enhancement via frequency-band extracting are investigated experimentally and theoretically in a chaotic optical-feedback semiconductor laser. At the optimal extracting bandwidth, the relationship between TDS and entropy evolution are investigated over the wide range of injection-feedback parameters. The chaotic dynamics is assessed by the noise-dominated threshold. We also experimentally and theoretically study the effects of extraction bandwidth and radio frequency on TDS and entropy growth. The experiment agrees well with the theory. The statistical distribution of chaotic laser intensity is improved at optimal extracting bandwidth, and the measured skewness of the intensity distribution is negligible. By use of this technique, desired chaotic sources can be prepared for physical random number generation and secure communication.

## 2. Experimental setup

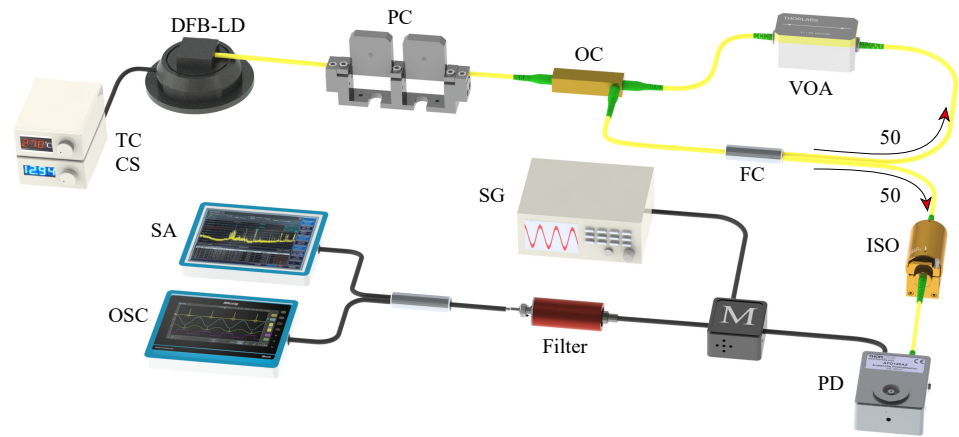

**Figure 1.** Experimental setup. TC, temperature controller; CS, current source; DFB-LD, distributed feedback laser diode; PC, polarization controller; OC, optical coupler; VOA, variable optical attenuator; FC, 50:50 fiber coupler; ISO, isolator; PD, photodetector; M, mixer; SG, signal generator; Filter, low-pass filter; SA, spectrum analyzer; OSC, oscilloscope.

Figure 1 depicts the diagram of the experimental setup. The chaotic laser is formed by a distributed feedback laser diode (DFB-LD) subject to optical feedback. The semiconductor laser with threshold current $J_{th} = 10.3$ mA is stabilized at 1550 nm by a low-noise current source (CS) with an accuracy of 0.1 mA and a temperature controller (TC) with an accuracy of 0.01 °C. The laser output passes through a polarization controller (PC), which serves to ensure polarization-maintained optical feedback. Then the light goes through an optical circulator (OC) and a 50:50 optical fiber coupler (FC). Half of the light goes back to the OC, and a digital variable optical attenuator (VOA) with a resolution of 0.01 dB is used to accurately control the feedback strength of the optical feedback loop. The feedback delay time is about 86.7 ns. The other half of the light enters a fiber optical isolators (ISO), which is used to ensure unidirectional transmission of chaotic laser. The final output is converted into an electrical signal by a fast photodetector (PD, bandwidth: 50 GHz). Then the signal is mixed with a radio-frequency (RF) signal generated by a signal generator (SG). A low-pass filter (LPF) with 3 GHz cutoff frequency (Mini-Circuits VLF-3000+) is used to extract the chaotic signal. The filtered signal is acquired by a spectrum analyzer (SA) with a bandwidth of 26.5 GHz and a real-time oscilloscope (OSC) with a bandwidth of 36 GHz and a sampling rate of 80 GS/s, respectively.

## 3. Theoretical model and methods
### 3.1. Lang-Kobayashi and frequency-band extraction model

A semiconductor laser with external optical feedback as a typical complex dynamical system can be well modeled by the Lang-Kobayashi (LK) equations [34]. The LK equations can also well reveal photon statistics of chaotic lasers and provide a necessary and better understanding of chaotic process [35]. The equations are expressed as follows:

$$\dot{E}(t) = \frac{1}{2}[G(t) - \tau_p^{-1}]E(t) + \kappa E(t - \tau_{ext})\cos[\phi(t)], \tag{1}$$

$$\dot{\varphi}(t) = \frac{\alpha}{2}[G(t) - \tau_p^{-1}] - \kappa E(t - \tau_{ext})E(t)^{-1}\sin[\phi(t)], \tag{2}$$

$$\dot{N}(t) = \frac{J}{e} - \frac{N(t)}{\tau_N} - G(t)|E(t)|^2, \tag{3}$$

$$\phi(t) = \omega\tau + \varphi(t) - \varphi(t - \tau_{ext}), \tag{4}$$

where $E(t)$ is the amplitude of complex electric field, $\varphi$ is the electric field phase, and $N$ is the carrier density. $\alpha$ is the linewidth-enhancement factor, $\tau_P$ is the photon lifetime, $\tau_N$ is the carrier lifetime. $\tau_{ext} = 2L/c$ is the external cavity delay time, $L$ is the length of the external cavity, $c$ is the speed of light. $\omega$ is the angular optical frequency, and $e$ is the electronic charge. The nonlinear optical gain is expressed as:

$$G(t) = G_N[N(t) - N_0]/(1 + \varepsilon|E(t)|^2), \tag{5}$$

where $G_N$ is the differential gain coefficient, $\varepsilon$ is the gain saturation factor, and $N_0$ is the carrier density at transparency. The optical feedback strength $\kappa$ is expressed as:

$$\kappa = (1 - r_{in}^2)r_0/(r_{in}\tau_{in}), \tag{6}$$

where $\tau_{in}$ denotes internal cavity round-trip time. In the simulation, the parameter values are set according to the laser operation in the experiment, as shown below: $\alpha = 5$, $G_N = 2.7 \times 10^{-8}\ ps^{-1}$, $N_0 = 1.36 \times 10^8$, $\tau_{ext} = 86.7\ ns$, $\tau_P = 3.2\ ps$, $\tau_N = 2.3\ ns$, $\omega = 1.216 \times 10^{15}\ rad/s$, $\varepsilon = 5 \times 10^{-7}$. The laser threshold current is $J_{th} = 10.3\ mA$, which is the same as that of the laser used in the experiment.

The original chaotic signals are acquired and improved by a frequency-band extraction, which can effectively suppress the TDS of high-dimensional chaotic laser signals. The theoretical model of frequency-band extraction is described as follows:

$$S(t) = LPF[I(t) \times cos(2\pi f_{LO}t)], \tag{7}$$

where $LPF[\cdot]$ represents a Chebyshev Type-II low pass filter with an effective bandwidth of 3 GHz, $I(t)$ is the intensity of original chaotic laser, $f_{LO}$ is the local RF frequency. By the frequency-band extraction, the chaotic signals with different bandwidth power spectra can be extracted depending on the frequency of the applied RF signal $f_{LO}$. It is worth noting that the original chaotic signals are mixed and heterodyned with a local RF frequency, and the heterodyned signals with redistributed broader power spectrum and suppressed lower TDS are produced. The high-frequency wide bandwidth is extracted and the low-frequency nuisances from TDS are effectively removed at the same time. To obtain high energy utilization around relaxation oscillation frequency and suppress the TDS simultaneously, we can optimize the $f_{LO}$ around relaxation oscillation frequency and choose the optimal extracting bandwidth (i.e. 3 GHz in this work). Obviously, the extracting bandwidth with the optimal energy utilization can be further improved as the original chaos bandwidth increases. For larger extracting bandwidth, we can increase the RF frequency to higher frequency region and choose wider bandwidth LPF. Meanwhile, the RF frequency is much higher than the LPF bandwidth. The large bandwidth chaos with high energy and low TDS contributes to the applications of chaotic random number generation and high-speed secure communication.

### 3.2. Time delay signature

The common approaches to quantify the TDS are the autocorrelation function, permutation entropy and delayed mutual information. In this work, we take the peak value of the autocorrelation function (ACF) at the optical feedback delay time to quantify the TDS. The ACF is defined as follows:

$$C(\Delta t) = \frac{\langle [I(t+\Delta t) - \langle I(t+\Delta t) \rangle][I(t) - \langle I(t) \rangle] \rangle}{\sqrt{\langle [I(t+\Delta t) - \langle I(t+\Delta t) \rangle]^2 \rangle \langle [I(t) - \langle I(t) \rangle]^2 \rangle}}, \qquad (8)$$

where $\Delta t$ denotes the delay time, $I(t)$ denotes the laser intensity and $\langle \cdot \rangle$ denotes time average. The peak value of the ACF at the external cavity delay time can be expressed as:

$$C_P = max|C(\Delta t)|_{\Delta t \in \lambda(\tau_{ext})}, \qquad (9)$$

where $\tau_{ext}$ denotes the external cavity delay time. The $C_p$ value provides useful information on the chaotic dynamics.

### 3.3. Entropy growth

PE is a robust complexity measurement based on the relative amplitude of time series values. Compared with Lyapunov exponent, strangeness of strange attractors, and Kolmogorov-Sinai entropy, the advantages of using PE as a metric are its implementation simplicity, flexibility, invariance, and robustness [29]. These advantages make PE especially applicable for analyzing reality-based analog signals generated by physical entropy sources. In this analysis, PE is obtained by constructing the probability distribution of ordinal patterns from the time series. The embedding dimension $d$ and embedding delay time $\tau$ are chosen appropriately. Embedding dimension $d$ is often recommended between 3 and 7 in practice. Due to time constraints, we choose $d = 4$ in this paper.

For a given time series $X = \{x_t, \ t = 1, \cdots, N\}$, we introduce the vector

$$s \to (x_{s-(d-1)\tau}, x_{s-(d-2)\tau}, \cdots, x_{s-\tau}, x_s). \qquad (10)$$

The ordinal pattern of the vector at the time ($s$) can be mapped into a unique permutation $\pi = (r_0, r_1, \cdots, r_{d-1})$ defined by

$$x_{s-r_0\tau} \geq x_{s-r_1\tau} \geq \cdots \geq x_{s-r_{d-2}\tau} \geq x_{s-r_{d-1}\tau}. \qquad (11)$$

The probability distribution $P(\pi_i)(i = 1, 2, \cdots, d!)$ is calculated by determining the relative frequency of all the $d!$ resulting permutations $\pi_i$.

The unnormalized PE is defined as

$$H_d = -\sum_n^{d!} P(\pi_i) \ln P(\pi_i). \qquad (12)$$

Then, we define the entropy growth $G_d$ as the difference of PE in neighboring embedding dimensions:

$$G_d = H_d - H_{d-1}. \qquad (13)$$

At the external optical-feedback delay time, the entropy growth $G_d$ not only reveals the entropy evolution of chaotic dynamics, but also has a relationship with the TDS.

In addition, it should be noted that the entropy growth $G_d$ can determine whether the chaotic dynamics is dominated by noise or not. There are noise-dominated and deterministic thresholds of entropy growth $G_d^{thr}$ used to assess the chaotic dynamics. When the $G_d$ is greater than the noise-dominated threshold $G_d^{Nthr}$, the $G_d$ increases with the embedding dimension $d$ and it indicates that the chaotic dynamics is dominated by noise. The noise limit of entropy growth is $G_d^{lim} = \ln d! - \ln(d-1)!$. For $d = 4$, the noise-dominated old of entropy growth $G_d^{Nthr}$ is 0.909. When the $G_d$ is less than the deterministic threshold $G_d^{Dthr}$, the $G_d$ decreases with the embedding dimension $d$ and it indicates that the dynamics is deterministic. For $d = 4$, the deterministic threshold of entropy growth $G_d^{Dthr}$ is 0.878. When the $G_d$ is almost constant, it means that the dynamics is in the intermediate regime.

## 4. Results

To experimentally evaluate the output signals of the chaotic laser, we first measure the power spectrum of original and extracted chaotic laser operating at bias current $J = 1.6J_{th}$ and feedback strength $\eta = 18\%$, as shown in Figure 2. According to the 80% bandwidth definition, the original spectrum of chaotic laser is about 8.0 GHz, as shown in Figure 2(a). The original chaotic signal is mixed down with a 4.2 GHz carrier and filtered by a LPF with 3 GHz cutoff frequency (Mini-Circuits VLF-3000+). Figure 2(b) shows the effective extraction bandwidth of 3 GHz accounting for 80% of the total 3.8 GHz bandwidth in the extracted chaos power spectrum. The extracted chaotic signal has a large clearance above the noise floor of 23 dB, which is higher than the power of most other frequencies. Moreover, the frequency of the external cavity $f_{ext}$ is about 0.01 GHz corresponding to the optical feedback delay time $\tau_{ext} = 86.7\ ns$, and the relaxation oscillation frequency $f_{RO}$ is around 4.2 GHz. It is worth noting that the extracted signal of chaotic laser avoids the external cavity period and high output energy around relaxation oscillation frequency is utilized sufficiently. Moreover, the extracting bandwidth with high energy utilization can be enhanced as the original bandwidth of chaotic laser increases. The power spectrum becomes flat and the chaotic dynamics evolves into the coherence-collapse regime where $f_{ext} \ll f_{RO}$.

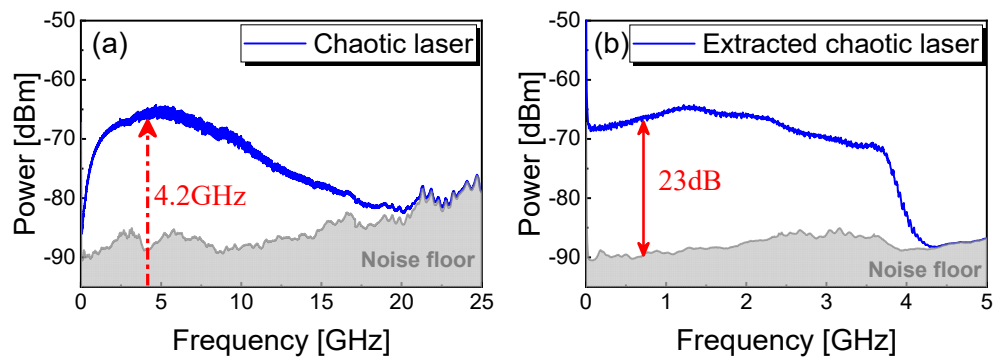

**Figure 2.** (a) Measured origin chaotic laser when $J = 1.6J_{th}$, $\eta = 18\%$ power spectrum; (b) extracted power spectrum of chaotic laser with 3 GHz effective bandwidth.

In order to further study the chaotic dynamics and entropy evolution, we experimentally measure the ACF and entropy growth, as shown in Figure 3. The parameters are the same as those used in Figure 2. In Figure 3(a), the peak value $C_p$ of ACF for chaotic laser is 0.442 at the optical feedback delay time of 86.7 ns. Compared to the original chaos, the TDS is suppressed up to 96% with the minimum of 0.016 via effective frequency-band extractor of 3 GHz. Moreover, the entropy growth $G_d$ ($d = 4$) at the feedback delay time is increased from 1.307 to 1.38 by the wideband frequency-band extracting, as shown in Figure 3(b). It should be noted that the 3 GHz frequency-band extraction significantly suppresses TDS of chaotic laser, and enhances the entropy growth $G_d$ that indicates the chaotic laser is in the noise-dominated regime ($G_d > G_d^{Nthr} = 0.909$).

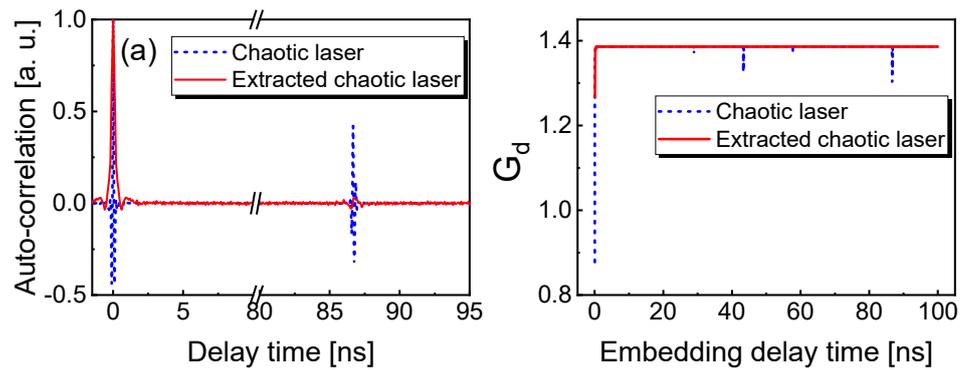

**Figure 3.** (a) ACF of origin chaotic laser and extracted chaotic laser. (b) Entropy growth $G_d$ of origin chaotic laser and extracted chaotic laser with embedding dimension $d = 4$.

Then we experimentally and theoretically investigate the influences of injection current and feedback strength on the chaotic dynamics and entropy growth, and characterize the relationship between the entropy growth $G_d$ and TDS. Figure 4 shows the influence of injection current on $C_p$ and $G_d$ through frequency-band extraction of 3 GHz effective bandwidth. At three feedback strengths $\kappa$ ($\eta$), the $C_p$ and $G_d$ show an inverse relationship for the injection currents. As the injection current increases, the TDS decreases first and then increases at weaker feedback strengths, while the variation trend of $G_d$ is inverse to that of TDS. At strong feedback strength $\kappa = 44\ ns^{-1}$ ($\eta = 35\%$), the TDS remains almost minimum and the $G_d$ becomes saturated for high injection currents. The theoretical results [Figure 4(a1)-4(c1)] are in good agreement with the experimental results [Figure 4(a2)-4(c2)].

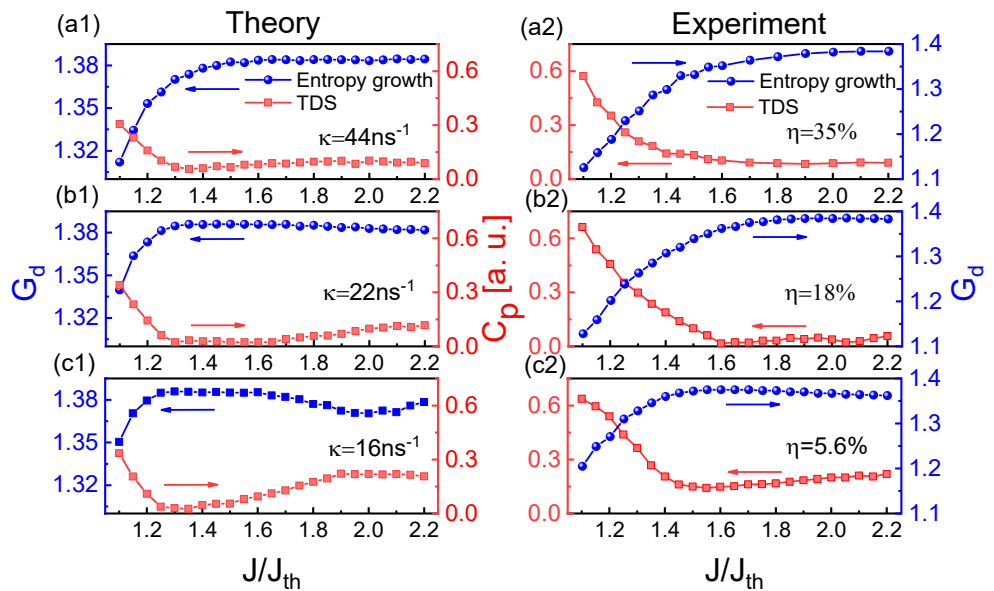

**Figure 4.** (a1)-(c1) Theoretical and (a2)-(c2) experimental results for $G_d$ and $C_p$ of extracted chaotic laser as a function of the injection current for $44\ ns^{-1}$ (35%), $22\ ns^{-1}$ (18%), $16\ ns^{-1}$ (5.6%). The embedding dimension $d$ is chosen as 4.

Figure 5 shows the influence of feedback strength on $C_p$ and $G_d$. The TDS $C_p$ and entropy growth $G_d$ also have an inverse relationship with the increase of the feedback strength. The experiment agrees well with the theory. At the injection current $J = 1.6J_{th}$, the TDS $C_p$ and entropy growth $G_d$ have more effective suppression and enhancement than those of other two injection-current cases. The above results also indicate that the evolutions of TDS and entropy growth with varying $J$ and $\kappa$ ($\eta$) are well revealed experimentally and theoretically. As a consequence, the suppression of TDS corresponds to the enhancement of entropy growth.

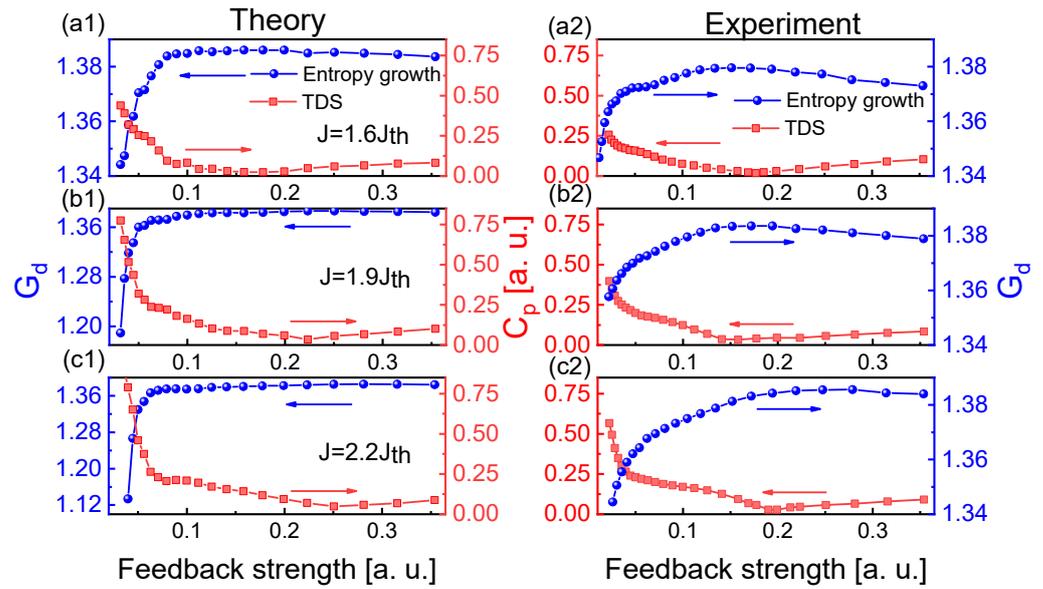

**Figure 5.** (a1)-(c1) Theoretical and (a2)-(c2) experimental results for $G_d$ and $C_p$ of extracted chaotic laser as a function of the feedback strength for $1.6J_{th}$, $1.9J_{th}$, $2.2J_{th}$. The embedding dimension $d$ is chosen as 4.

For clarity, we also study the influences of RF frequency and extracting bandwidth on the TDS suppression of the extracted chaotic laser. Figure 6 shows the experimental and theoretical TDS variations of extracted chaotic laser under different RFs. Except the RF, the operating parameters of the extracted chaotic laser are the same as those used in Figure 2. Figure 6(a) shows the $C_p$ values of TDS versus the RF with the extracting bandwidth of 580 MHz. When the RF is around 1.18 GHz with the 580 MHz extracting bandwidth, the maximum $C_p^{max}$ and the minimum $C_p^{min}$ are 0.2843 and 0.0308, respectively. Figure 6(b) shows the maximum value $C_p^{max}$ of TDS is 0.266 and the minimum $C_p^{min}$ is 0.058 when RF is around 2.18 GHz with extracting bandwidth of 1 GHz. Figure 6(c) shows the maximum and minimum values of $C_p$ are 0.4842 and 0.016 when the RF is around 4.2 GHz with extracting bandwidth of 3 GHz. The reason for the increase of the maximum $C_p^{max}$ is that the RF frequency is located at the relaxation oscillation frequency of Figure 2(a). The effective suppression of TDS shows periodic variations and the minima $C_p^{min}$ appear in every $f_{ext}/2 = 6$ MHz. The periodic peaks and valleys of $C_p$ correspond to $(2n + 1) f_{ext}/4$ and $n f_{ext}/2$, respectively. Meanwhile, we adopt the LK and frequency-band extraction model to verify the experimental results, and the solid curves indicate the theoretical fitting in Figure 6.

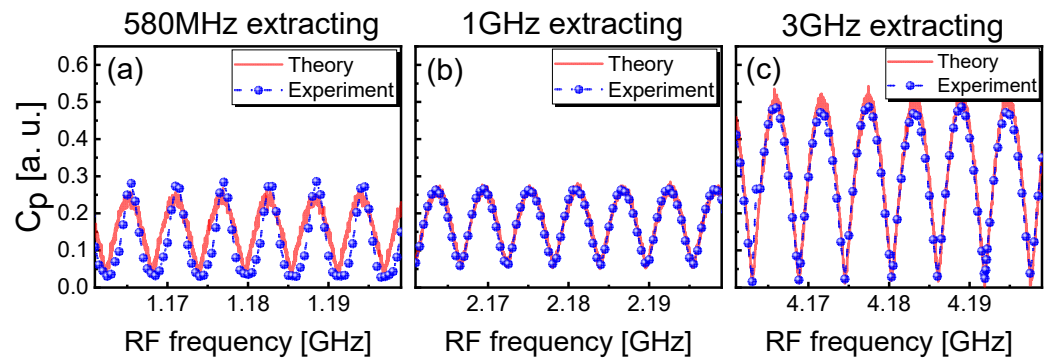

**Figure 6.** Theoretical and experimental TDS results of the extracted chaotic laser versus RF frequency for (a) 580 MHz, (b) 1 GHz, (c) 3 GHz extracting bandwidth.

The LPF bandwidth affects the minima of $C_p$ in the TDS suppression and the corresponding entropy growth $G_d$ of the extracted chaotic laser. In Figure 7, the theoretical and experimental $C_p^{\min}$ and $G_d$ of the extracted chaotic laser are shown for various LPF bandwidths. The extracted chaotic laser operates at the same parameters as those used in Figure 3, except the LPF bandwidth. The RF in frequency-band extraction is around 4.2 GHz, which is optimized to fully utilize the high energy and broaden the power spectrum around the relaxation oscillation period. It is noted that the TDS is suppressed effectively and the minima $C_p^{\min}$ remain almost unchanged for different extracting bandwidths. The inverse relationship is between the minima $C_p^{\min}$ and entropy growth $G_d$.

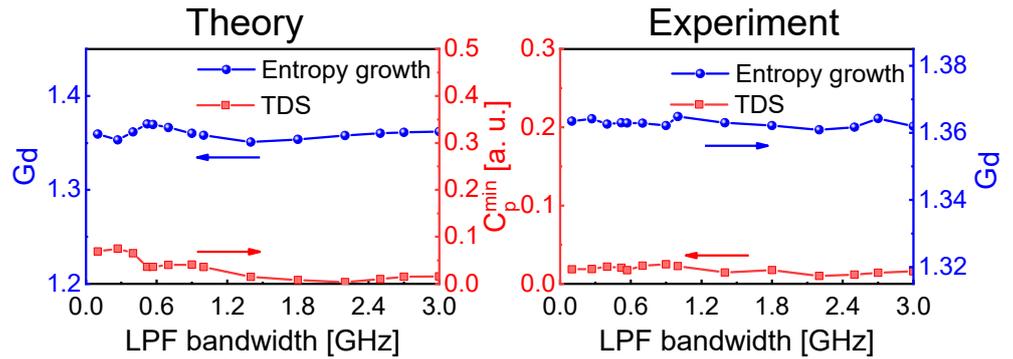

**Figure 7.** (a) Theoretical and (b) experimental results of the minima $C_p^{\min}$ in the TDS suppression and the entropy growth $G_d$ of extracted chaotic laser as a function of LPF bandwidth.

The frequency-band extraction not only effectively suppresses the TDS of chaotic laser, but also well improves the statistical distribution of the chaotic intensity. Figure 8(a) shows the measured time series and its probability statistical distribution of the original chaotic intensity. The parameters of the original chaotic laser are the same as those used in Figure 3. The statistical distribution of the chaotic laser intensity obviously deviates from the Gaussian distribution, due to the disturbances of external cavity and relaxation oscillation periods. The deviation is mainly from the non-Gaussian or deterministic signals in the dynamical process. The distribution skewness of the original chaotic intensity is 0.577. The red solid curve represents the fitting of Gaussian random distribution, and the green dashed line represents the normalized statistical mean of the measured intensity. Figure 8(b) shows the measured intensity and its probability statistical distribution through 3 GHz frequency-band extraction. The extracting parameters are the same as those used in Figure 3. In this case, the probability statistical distribution of the extracted chaotic intensity is improved significantly, which agrees well with Gaussian random distribution. The experimental skewness of the extracted intensity distribution is 0.001. Moreover, the distribution skewness of chaotic laser intensity as a function of the feedback strength is further investigated experimentally, as shown in Figure 8(c). With the increase of feedback strength, the distribution skewness of the original chaotic intensity decreases first and then increases, which is much greater than 0. It is indicated that the intensity distribution of the original chaotic laser obviously deviates from Gaussian random distribution. With 3 GHz frequency-band extraction, the measured skewness of the intensity distribution decreases by more than two orders of magnitude. The minimum skewness remains almost unchanged for different feedback strengths. That means wide frequency-band extraction effectively improves the intensity statistical distribution and randomness of the chaotic laser.

## 5. Conclusions

In summary, we theoretically and experimentally investigate the TDS suppression and entropy growth enhancement of an optical-feedback chaotic laser via frequency-band extracting. By 3 GHz effective bandwidth extracting, the TDS suppression ratio is reached to 96% with the minimum $C_p^{\min}$ of 0.016 compared to the original chaotic laser. The inverse relationship between the TDS and entropy growth is verified theoretically and experimentally over a wide range of the injection currents and feedback strengths. The influences of RF frequency and extracting bandwidth on the TDS and entropy growth are revealed theoretically and experimentally. By optimizing the RF frequency and extracting bandwidth of frequency-band extractor, the TDS suppression and entropy growth enhancement are simultaneously achieved with wide frequency-band extraction. It is also exploited to efficiently utilize the high-energy output around the relaxation oscillation frequency, broaden the chaos bandwidth and meliorate power spectrum flatness of the chaotic laser. The experiment is in good agreement with the theory. Furthermore, the wide frequency-band extraction effectively improves the probability statistical distribution of chaotic intensity, and the measured skewness of intensity distribution approaches 0 for various feedback strengths. The above improvements of chaotic laser via the frequency-band extraction are beneficial to chaos-based communication and physical random number generation.

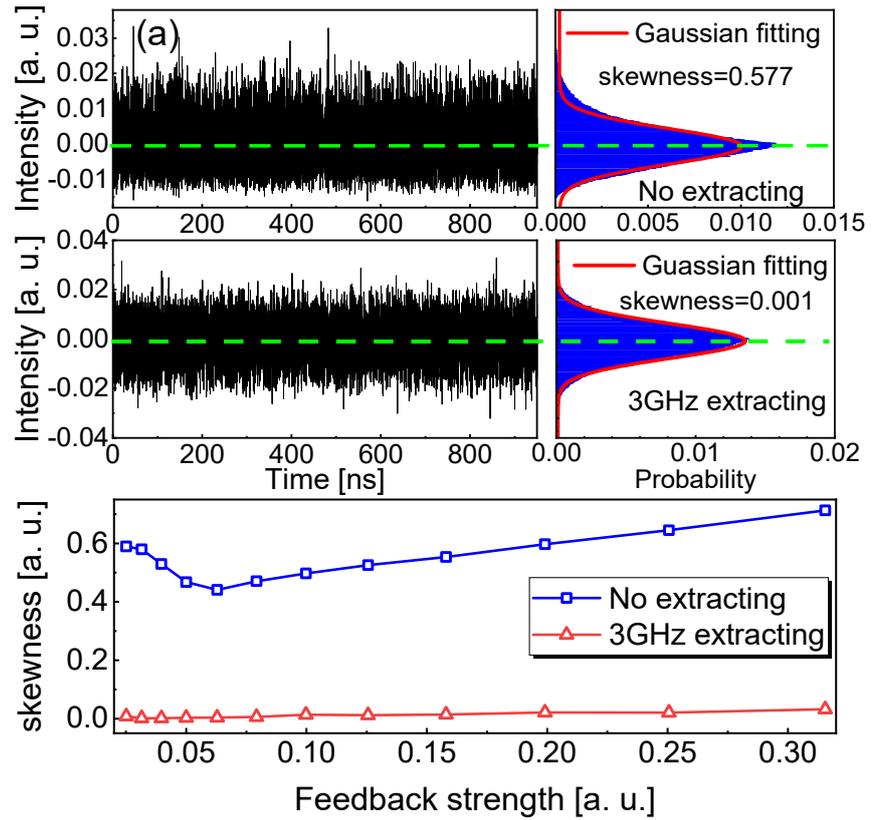

**Figure 8.** Experimental results of time series, intensity distribution and its skewness of the chaotic laser (a) before and (b) after 3 GHz frequency-band extraction. (c) measured skewness of intensity distribution versus feedback strength at the injection current $J = 1.6J_{th}$.

**Author Contributions:** Conceptualization, Y.G. and X.G.; methodology, Y.G. and T.L.; software, T.L. and H.Z.; validation, T.L., T.Z. and H.Z.; supervision, X.G.; writing—original draft preparation, Y.G. and X.G.; writing—review and editing, Y.G. and X.G. All authors have read and agreed to the published version of the manuscript.

**Funding:** This research was funded by National Natural Science Foundation of China (Grant Nos. 61875147, 62075154 and 61731014), Key Research and Development Program of Shanxi Province (International Cooperation, 201903D421049), Shanxi Scholarship Council of China (Grant No. HGKY2019023), Natural Science Foundation of Shanxi Province (Grant No. 201801D221182), Scientific and Technological Innovation Programs of Higher Education Institutions in Shanxi (Grant Nos. 201802053, 2019L0131) and Program of State Key Laboratory of Quantum Optics and Quantum Optics Devices (Grant No. KF201905).

**Acknowledgments:** The authors would like to thank Martin Virte and Tiancai Zhang for fruitful discussions.

**Conflicts of Interest:** The authors declare no conflict of interest.